\documentclass[a4paper,11pt]{article}
\usepackage{authblk}
\usepackage[utf8]{inputenc}
\usepackage{fullpage}
\usepackage[OT4]{fontenc}
\usepackage{hyperref}
\usepackage{todonotes}
\usepackage{amsmath}
\usepackage{subfigure}
\usepackage{microtype}
\usepackage{setspace} 

\usepackage{algorithmic}
\usepackage{algorithm}
\newcommand{\procName}{Procedure}
\floatname{algorithm}{\procName}

\title{Statistical analysis of geoinformation data for increasing railway safety}

\author[1]{Katarzyna Gawlak}
\author[1]{Jarosław Konieczny}
\affil[1]{{Department of Railway Transport, Silesian University of Technology},
		{Krasinskiego~8},
		{40-019},
		{Katowice},
		{Poland}}
\author[2]{Krzysztof Domino}
\author[2]{Jarosław Adam Miszczak}
\affil[2]{{Institute of Theoretical and Applied Informatics, Polish Academy of Sciences}, {Baltycka~5}, {44-100}, {Gliwcie},{Poland}}

\begin{document}

\maketitle

\begin{abstract}
The impact of rail transport on the environment is one of the crucial factors for the sustainable development of this form of mass transport. 
We present a data-driven analysis of wild animal railway accidents in the region of southern Poland, a step to create the train driver warning system.
We built our method by harnessing the Bayesian approach to the statistical analysis of information about the geolocation of the accidents. 
The implementation of the proposed model does not require advanced knowledge of data mining and can be applied even in less developed railway systems with small IT support. Furthermore, we have discovered unusual patterns of accidents while considering the number of trains and their speed and time at particular geographical locations of the railway network. We test the developed approach using data from southern Poland, compromising wildlife habitats and one of the most urbanised regions in Central Europe, based on this we conclude that our model is best suited to railway lines that pass through varying types of landscape.
\end{abstract}
  
\section*{Keyword}
Bayesian analysis; geographic data analysis; mass transport safety; wide life ecology

\section*{Highlights}
\begin{itemize}
\item geographical statistics of train accidents with wild animals were analysed,
\item easily implementable Bayesian model of accident prediction was created,
\item the utility of the predictive model was demonstrated on real data,
\item issues of wildlife ecology, and resilient and safe rail transport were addressed,
\item unusual patterns of accidents involving wild animals were recorded.
\end{itemize}

\section{Introduction}

The history of train collisions with wildlife is as old as the railway itself, and with the growing network of railway connections and the increasing speed of rolling stock vehicles, the problem is expected to grow~\cite{Animalscollisions_CzechRepublic}. It is not without significance that railway lines very often cross the natural habitats of wild animals. 
Additionally, wild animals do not lead a sedentary lifestyle and migration is part of their existence. 

The demand for passenger transport is related to the quality of service and, in particular, to its resilience. It is very easy to lose passengers due to divergence from the planned schedule. Even a delay of several minutes causes some dissatisfaction among passengers, see Fig.~5 of~\cite{MONSUUR202119}. Methodologically, disturbances in public transport may be divided into different categories according to dimensions related to planning, probability, impact, time of occurrence, duration time, scope of location, frequency, and miscellaneous referring to the up-to-date review of disturbances in passenger traffic~\cite{ge2022robustness}. 

Collisions with wild animals are important examples of disturbances in passenger traffic that affect passenger satisfaction, particularly in the context of the competitiveness of public transport compared to private means of transport~\cite{Passengers_satisfaction}. What is also important here is the fact that collisions with wild animals cause disturbances that propagate along the rail network, causing passenger dissatisfaction at various locations. This is not a desired situation, as the railways provide one of the most promising means of transportation from an environmental point of view. Moreover, reducing the number of collisions with wild animals has a positive effect on the environment itself, saving animals and the local heritage tied to them. 

Aside from the passenger satisfaction issue, such collisions also have a negative impact on railway operation~\cite{Collisions_cost_risk}. This includes the costs of the repair of the rolling stock, cleaning, etc. Besides, it is important to note that the construction of new types of rolling stock is more vulnerable to collision with animals than the old ones~\cite{Railway_ecology}. Furthermore, currently, trains move faster, but they are quieter~\cite{Brakes_noise} than in past years. This may reduce the effect of scaring away animals, especially in combination with particular landscape formations~\cite{BACKS2022100516}.
The overall socio-economic costs of train collisions with wildlife in Sweden have recently been estimated at 100,000,000–150,000,000 Euros per year \cite{seiler2014costs}.
This is similar to the costs estimated for wildlife-vehicle collisions on roads (250,000,000 Euros per year) \cite{seiler2017wildlife}, even though railways comprise less than 2\% of the national road network in Sweden. Such an observation provides the sound motivation for this research. Concerning Poland and referring to the Intercity operator (PKP Intercity, eng. Polish State Railways Intercity), the cost of repair of rolling stock after collisions with wild animals in 2014 was estimated at the value above 100,000,000 Euros \cite{collisions_Konieczny}.

The goal of this work is to introduce the proactive method of improving resilience in rail transport and mitigate its negative impact on the environment. In particular, we concentrate on disturbances tied to collisions with wild animals. As a resilient railway system, we understand the ability of such a system to provide effective and satisfactory (for passengers) services regardless of problems with the usual functions of the system. Therefore, the system is expected to accommodate and recover quickly from disruptions or disasters~\cite{doi:10.1080/01441647.2020.1728419}. From the experience of our partner, the local railway operator in Silesia (south part of Poland), usually collision with a wild animal causes a delay of the train of several minutes, which is related to the obligation to perform a visual inspection of the vehicle by the train driver. As mentioned above, such disturbances would propagate along the network, causing other disturbances. This would affect the resilience of the above-mentioned railway system. Besides, more severe cases also appear sporadically and the vehicle may be ordered to move to the depot due to technical problems caused by the accident.

Our unique contribution is a complete approach, starting from data on collisions of animals, including the railway traffic density, and ending on the predictive and warning mechanism for train divers. We built our method using information about the geolocation of the accidents and harnessing the Bayesian approach. Hence, we avoid the complexity of methods involving parametric models of animal activity. Because of this, even if the analysis is performed for particular data from south Poland, the general methods and tools we present can also be applied to different types of data from different regions of the world.

This paper is organised as follows. In Section~\ref{sec:preliminaries} we provide the literature review of the analysis of collisions of trains with wild animals and outline the elements specific to the data in the analysed region. In Section~\ref{sec:algorithm} we introduce the data-driven algorithm, based on the statistical analysis of incident frequencies following the Bayesian approach. In Section~\ref{sec:data_analysis} we perform a series of tests on the supplied data. Finally, in Section~\ref{sec:discussion} we summarise the results presented.

\section{Preliminaries}\label{sec:preliminaries}

Before introducing the proposed approach, we provide some background information important for the presented methods. We start with a review of the predictive models used in the literature concerning the animal-train collisions. We argue that there are some important elements missing in the existing approaches. The goal of our work is to address these shortcomings.

Additionally, we include a discussion of the elements which are specific to the case elaborated in this work. Even if our approach can be adapted to any data set, we believe that it is important to clarify some nuances of the analysed data set to better understand the model proposed in Section~\ref{sec:algorithm}. 

\subsection{Review of predictive models}\label{sec:review}

In the literature, many elaborate models of patterns of collisions with trains and animals have been discussed. However, we have not found a particular applicability of these models in terms of improving the resilience of rail transport. The intensity of railway traffic in comparison with accidents has also not been particularly investigated (i.e., is the hot-spot the place where we have very dense railway traffic and numerous accidents). 
We have found no wide analysis considering the number of trains passing particular lines tied to accidents.

Starting from~\cite{malo2004can} predictive models have full meaning applications in the prediction of collisions with wild animals. The problem of trains colliding with wild animals appears all over the world~\cite{Santos2017}. The specifics of this problem are related to particular regions and spaces of animals leaving there; compare Asia \cite{Roy2017} \cite{article}, North America~\cite{Heske2015BloodOT,St.-Clair:2020aa}, South America \cite{Brazil}, and Europe \cite{Godinho2017,ruben2012railroad,batsinPoland}. There can be drown the general conclusion from the analysis of such events -- there appear hot-spots -- places where the intensity of collisions is high. The challenge is to localise spatial and temporal coordinates in such places. This is the first step to undertake countermeasures to mitigate disturbances due to such accidents. In particular, these accidents are usually analysed with regard to the density of animals in the population, as well as the density of railway traffic. This leads to the spatial patterns approach, used in various analysis across the world. A prominent example of this approach is the analysis of accidents with elephants in India~\cite{ahmed2022pandora}. Another model demonstrates the link between kangaroo-train collisions, train speed, and the coincidence of periods of high train and kangaroo activity in Australia \cite{visintin2018managing}.

Let us now concentrate on Europe, the region for which our analysis has been performed. The spatial patterns approach has been applied to analyse the collisions of train with moose (\emph{Alces alces}) \cite{moose-Norway} in Norway. In a similar priori mentioned, hot-spots of animal collisions have been located in Romania and Sweden~\cite{seiler2017wildlife,land10070737}.
Our analysis was performed for the south region of Poland, that is, close to Czechia, for which the analysis of the hot-spots has been identified using the KDE + (Kernel Density Estimation +) method~\cite{Animalscollisions_CzechRepublic}. Mentioned KDE+ extends the standard KDE (Kernel Density Estimation) by adding repeated random simulations using the Monte Carlo approach \cite{bil2013identification}. In particular, the impact of various parameters on collisions has been analysed for the Czech railway network.
The neighbourhood of the railway line has been parameterised, e.g. by the distance to the closest forest area, the presence of water reservoirs, and urban areas. Further parameters of the railway line have been considered, including maximum speed, presence of overhead electric wires, and track geometry~\cite{Poster_animals_CZ}.

\begin{table}[]
\begin{tabular}{|l|l|l|l|l|}
\hline
article         & animal        & region    & \begin{tabular}[c]{@{}l@{}}use external data \\ (landscape)\end{tabular} & \begin{tabular}[c]{@{}l@{}}predictive  method\end{tabular}    \\ \hline
\cite{clauzel2017evaluating} & \begin{tabular}[c]{@{}l@{}}Tree frog\end{tabular}  & \begin{tabular}[c]{@{}l@{}}France\end{tabular} & Yes (habitats)   & \begin{tabular}[c]{@{}l@{}}landscape \\ graph construction\end{tabular}         \\ \hline
\cite{Roy2017}     & Elephant     & India     & Yes (eleph. signs)    & \begin{tabular}[c]{@{}l@{}}Kernel/ point density\end{tabular}  \\ \hline
\cite{St.-Clair:2020aa}      & \begin{tabular}[c]{@{}l@{}}Bear, carnivore\\ ungulate\end{tabular} & Canada    & \begin{tabular}[c]{@{}l@{}} Yes   (proximity to water, \\ amount of water)   \end{tabular}   & \begin{tabular}[c]{@{}l@{}}  Regression \\ multi parameters \end{tabular}         \\ \hline
  \cite{ruben2012railroad}  & Tortoise &  Romania & Yes (habitats)  & \begin{tabular}[c]{@{}l@{}}  Factor model \\ tied to habitants \end{tabular}     \\ \hline
    \cite{batsinPoland}  & Bat &  Poland & \begin{tabular}[c]{@{}l@{}} Yes (bat  routine \\ and habitats observation) \end{tabular}  & \begin{tabular}[c]{@{}l@{}}  General linear mixed \\ model (GLMM)  \end{tabular}     \\ \hline
    \cite{ahmed2022pandora}  & Elephant &  India & \begin{tabular}[c]{@{}l@{}} Yes (spatial relationship \\ elephant density) \end{tabular}  & \begin{tabular}[c]{@{}l@{}}  Geographically \\ weighted regression  \end{tabular}     \\ \hline
    \cite{visintin2018managing}  & Kangaroo &  Australia & \begin{tabular}[c]{@{}l@{}} Yes (kangaroo occurrence) \end{tabular}  & \begin{tabular}[c]{@{}l@{}}  Regression model  \end{tabular}     \\ \hline
    \cite{moose-Norway} & Moose & Norway & Yes (vegetation removal)& Cost-benefit analysis \\ \hline 
    \cite{seiler2017wildlife}& \begin{tabular}[c]{@{}l@{}}Moose, roe deer, \\ reindeer\end{tabular} & Sweden & \begin{tabular}[c]{@{}l@{}} Yes (composition of the \\ surrounding landscape, \\ availability of forage, \\ preferred habitat) \end{tabular} & Logistic regression \\ \hline
    \cite{land10070737} & \begin{tabular}[c]{@{}l@{}} Brown bear, lynx, \\ wolf, red deer, \\ roe deer, wild boar \end{tabular} & Romania & \begin{tabular}[c]{@{}l@{}} Yes (number of \\ individual at location) \end{tabular} & \begin{tabular}[c]{@{}l@{}} Weighted factors \\ analysis \end{tabular} \\ \hline
        \cite{Animalscollisions_CzechRepublic}  & \begin{tabular}[c]{@{}l@{}}Roe deer, wild boar,\\ red deer, red fox,\\ fallow deer, European\\ badger, European hare,\\ muflon\end{tabular}  &  Czechia & \begin{tabular}[c]{@{}l@{}} No \end{tabular}  & \begin{tabular}[c]{@{}l@{}}  Kernel density \\ + Monte Carlo \\ method derived for \\ animals cars collisions \end{tabular}     \\ \hline
        \cite{Poland-analysis} & \begin{tabular}[c]{@{}l@{}} Roe deer, red deer,\\ moose, wild boar \end{tabular} & Poland & No &  \begin{tabular}[c]{@{}l@{}}Generalised additive \\ mixed models \\ (GAMMs) uses only \\ temporal data \end{tabular} \\ \hline
        \cite{Wielkopolska}& \begin{tabular}[c]{@{}l@{}} Roe deer, wild boar,\\ red deer\end{tabular} & Poland & \begin{tabular}[c]{@{}l@{}}Yes (presence of specific \\land use function)\end{tabular} & \begin{tabular}[c]{@{}l@{}}Spatial analyses\\ (QGIS, QChainge) 
    \end{tabular} \\ \hline
    \end{tabular}
\caption{Summarising and comparison of relevant studies on predictive models of rail animals collisions. Observe, that there are only two works that do not consider external data from the landscape.}
\label{tab::existing_methods}
\end{table}

In Poland, analysis of train collisions with wild animals for the time period 2012 - 2015 supplied by the national network operator (PKP PLK S.A.) was discussed in \cite{Poland-analysis}. 
There, the temporal (day-time and year-time) distribution of collisions has been analysed and compared with the patterns of the wildlife of various animals. For data analysis, generalised additive mixed models (GAMMs) were applied. Spatial analyses with open-source Geographic Information System (GIS) software were performed for chosen railway lines in the Greater Poland region, in particular on railway lines close to Poznań for 10 years (i.e., 2007 - 2017). In particular, spaces such as wild boar, red deer, roe deer, and lands such as forests, waters, developed areas, and agricultural land were analysed. The conclusion was that the analysis of collisions of trains with wild animals has to be performed separately for each space, as their behaviour is different for various lands~\cite{Wielkopolska}.

From the introduction and literature review we can conclude, that the analysis of temporal and spatial localisation of collisions with wild animals has been performed in Poland and other countries around the world, also using GIS tools. The results of the analysis demonstrate that there appears hot-spots, i.e. places where collisions are frequent. Furthermore, statistical analysis of basic features, such as time, particular spice ecological environment, speed of trains, etc. has been performed. This was tied to the possible implementation and analysis of acoustic animal warning devices.

he relevant studies on predictive models of train-animals collisions are summarised in Table~\ref{tab::existing_methods}. From this summary, one can conclude that most studies input external data (called also landscape data) such as landscape form, animal paths, animal habitat, etc. These data are problematic to acquire by the railway operator, making such a model less practical. We have found two works that do not consider such external data, the first~\cite{Animalscollisions_CzechRepublic} uses a Kernel density + Monte Carlo method that is more dedicated to road-animals collisions and uses only limited temporal relation of data, the second method~\cite{Poland-analysis} does not consider spatial relation of data, henceforth it cannot be applied in hot-spot detection. We recognise the scientific gap of the lack of systematic research on railway-dedicated collision hot-spot detection, given statistical data that the railway operators can easily acquire.

\subsection{Case-specific factors}\label{sec:case-specific}

One should note that in Poland, the driver is obliged to stop the train immediately after the collision with the obstacle. Then the technical check of the rolling stock has to be performed. Frequently encountered faults are the violation of the vehicle sheathing, vehicle scrapers, damage to the sandblaster's wires, and in more serious cases, which may already have a significant impact on safety, these are failures of the braking systems, which may even result in the need to tow the vehicle. All of these problems add up to the delay and disturbance of the train.
It is important to note that the driver has a range of tools that can be used to prevent a collision. These are: ringing the horn or applying emergency brakes~\cite{BHARDWAJ2022114992}, but there also exist some vehicle-based warning systems~\cite{Train_system}. Hence, we believe that automatic warning before entering the zone where collisions with wild animals are possible would be the important tool in mitigating the problem of such collisions. To find such zones in time and space, we propose the systematic approach, based on the analysis of the locations and times of events. 

The specifics of the railway traffic (in contrast to the road traffic) yield less continuous flow of vehicles; i.e., the railway traffic is an interval-like one with time windows, where animals can safely pass the track~\cite{Railway_ecology,railway_vs_road_ecology}. This observation has been confirmed by the research conducted in Poland on the UOZ (animal protection device acoustic deflector) device~\cite{UOZ-1device}. In detail, out of $2262$ animals recorded near the track $76\%$ appeared in the time window, where no trains were present.

There are two types of warning devices that are dedicated to reducing collisions with wild animals. These are devices located along the railway line or devices placed directly on the vehicle. These can be classified as infrastructure-based countermeasures and trains-based countermeasures. The first type consists of two routines, i.e. detection of the vehicle and warning away animals~\cite{BACKS2017563}. In Poland, the UOZ device is integrated with the signalling system and emits natural animal warning calls~\cite{UOZ-1_producent}. There is also an option for the light signal, i.e. warning reflectors~\cite{Warining_reflectors_Poland}. In Poland, these systems are in the domain of the network infrastructure manager. The second type, on devices placed on vehicles, has been applied, for example, in Japan~\cite{Train_system}.

In the network analysed in this paper, there are no warning systems described above, and our analysis is performed from the perspective of the railway operator, that is, independent of the infrastructure manager. The interest of the operator is to have a dedicated system to support drivers in the detection of animals for the adequate prevention. Henceforth, we opt for the algorithmic support system for divers. An example of these is the approach using thermovision cameras~\cite{BHARDWAJ2022114992}. This approach has some difficulties with respect to the proper determination of the distance to animals and the detection of animals in areas of intense vegetation. Referring to the above, in our opinion, there are no reliable and generally applicable tools for warning train drivers of approaching areas with the risk of collisions with wild animals. The aim of the presented study is to develop a methodology for linking the probability of collision with parameters such as the spatial location of the train, the time of day and the season of the year.

\section{Bayesian model of warning }\label{sec:algorithm}

In this section, we provide the details of the Bayesian model of marking line segments with warning. The model is based on past data on train collisions with animals.  We start with an initial discussion of the underlying assumptions. Next, we provide a detailed description of the procedure, which can be implemented using any system that supports elementary statistical procedures.

\subsection{Initial considerations}

The mechanism of animals-trains interaction is complex, and many parameters have to be considered to make these models suitable for analysis,  In particular. It depends on spatial and temporal parameters (e.g. the particular daytime). In detail, one has to take into account, for example, the chance of an animal leaving the track in the time window starting when the train is noted by the animal and ending when it is too late to escape from the collision. Here, if the train is not observed early enough by the animal, the chance of an accident increases. However, the uncontrolled behaviour of an animal, including panic, may affect the above mentioned chance in an unexpected way. Furthermore, the train that is far away or slowly moving, may be observed by the animal just as the ``large object'' that poses no danger. This factor may suggest some unusual negative relation between a train speed and the probability of collision, and we include discussion of this issue in the further part of this work.

For detailed analysis of some of these factors, in~\cite{Reakcja_zwierzyny} the reaction of the animals was analysed with reference to the light and sound signals of the approaching train. In particular, the so-called flight initiation distance (FID). Basically, this reaction is different for different types of animals. Interesting that the train's horn may have an unusual effect on animals. e.g. in the case of roe deer, its runaway starts, respectively, early because of the horn, but may be directed toward the danger (the train). Importantly, for our perspective, there is the full meaning impact on accidents of the particular time with reference to the daylight cycle. The daylight has a non-trivial impact on the visibility of the animal and the field of view of both the animal and the driver
~\cite{BHARDWAJ2022114992}.

The complexity of the behavioural pattern and the plethora of factors which have to be taken into account for creating the analytical model of the animal behaviour suggest that this approach is not feasible in most cases. This is especially true as some of the factors might be very hard to measure, and thus the resulting model might be heavily biased. On the other hand, any parametrial model developed using set of parameters from one location will be unsuitable for use in other locations and situations. Thus, in the proposed approach, we built a system to detect parts of railway lines with an increased probability of accident using a Bayesian approach. 

\begin{figure}[ht!]
\centering
\includegraphics[width=\textwidth]{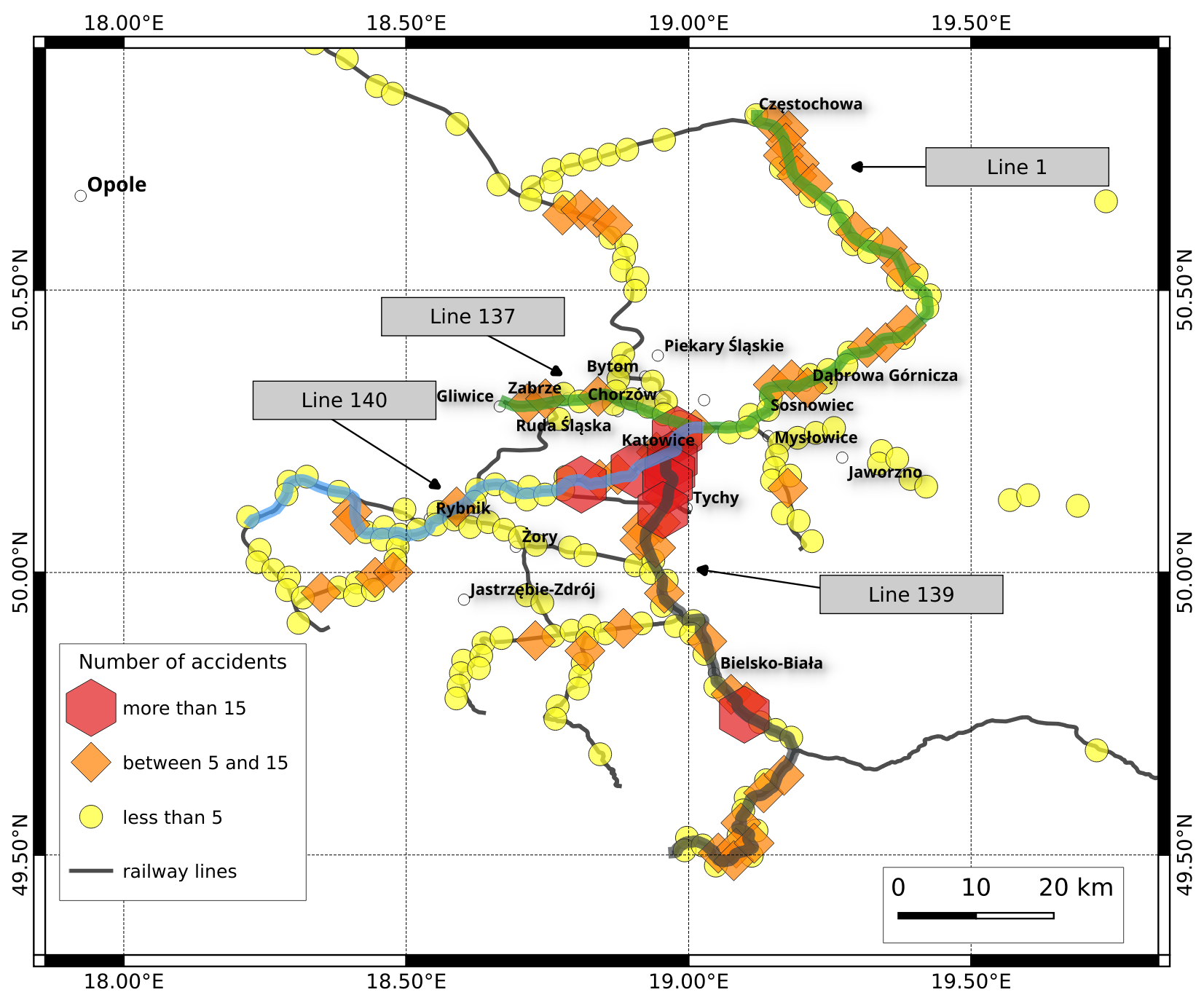}
\caption{Location of areas according to the number of all accidents during the period of three years 2020-2022. The raw data about accidents is processed to assign the number of accidents to the locations. Markers are calculated by counting the number of points in the hexagonal grid with 2.5 km vertical and horizontal spacing (created using QGIS \texttt{Create grid} processing tool). The most important lines are marked in colours.} 
\label{fig:locations}
\end{figure}

As a starting point in our approach, we use data on accidents with detailed information about the species involved, the time, and the location of the accident. The map of accidents concerning local railway operators in Silesia in 2021 is presented in Fig.~\ref{fig:locations}. This is in some analogy to \cite{Roy2017} where the spatial distribution of train accidents with animals has been demonstrated in order to determine hot-spots.

\begin{figure}[ht!]
\subfigure[hourly profiles for roe deer]{
\includegraphics[width=0.45\textwidth]{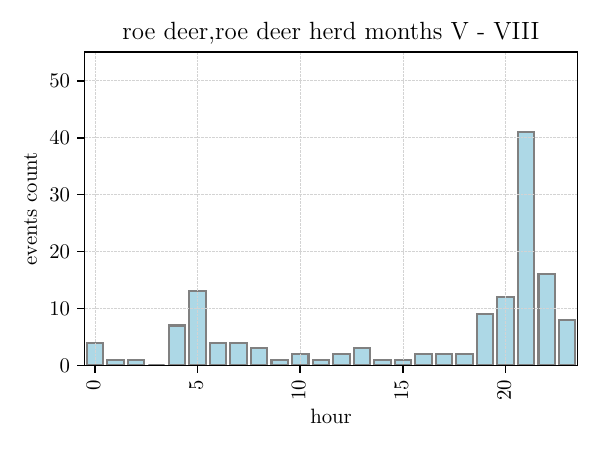}
\includegraphics[width=0.45\textwidth]{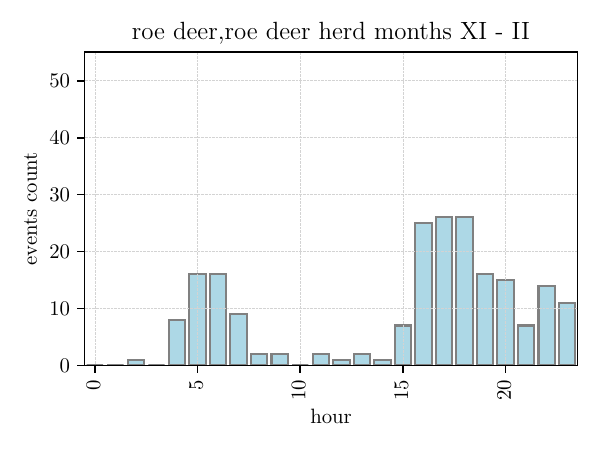}
}
\centering
\subfigure[species profiles]{
\includegraphics[width=0.5\textwidth]{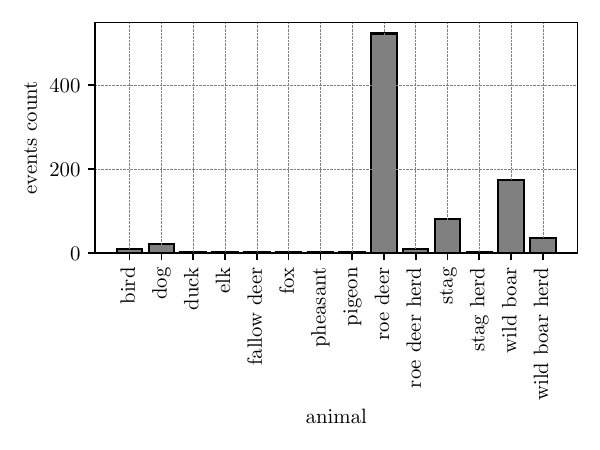}}
\caption{Profiles for animal-train collisions for various species. Spices of animals that took part in accidents, and seasons profiles of accidents with roe deer.}
\label{fig:various_profiles}
\end{figure}

Please note the spatial differences in the number of accidents. Based on the above discussion, we plan to take into consideration-specific profiles of accidents regarding particular animals, as well as daily profiles given various parts of the year. See Fig.~\ref{fig:various_profiles} (upper panel) for various profiles of accidents with roe deer depending on the part of the year.

\subsection{Model details}
The goal of our model is to take into consideration the spatial and temporal assessment of accidents. For this reason, we follow the recent approaches to train accidents with wild animals~\cite{ahmed2022pandora,visintin2018managing}. In \cite{ahmed2022pandora} train collisions with elephants for years 1990-2018 in Indian railways were analysed. The spatial and temporal profiles of the elephant wildlife reservoirs were then estimated. In~\cite{visintin2018managing} the seven-year period 2009-2015 of data with collisions of trains with kangaroos was examined. There, the particular model of the temporal and spatial activity pattern of kangaroos is developed. 
Following these two papers, we also develop the risk model and predict the collision risk. However, in contrast to the models proposed in~\cite{ahmed2022pandora} and \cite{visintin2018managing}, our method is dedicated to the case where no particular data on activity patterns of animals (e.g. reservoirs, migrations) are available. Hence, we estimate all parameters of the model from the raw data of train animal collisions.

Namely, we estimate from past data the probability that for a given day and given train, the accident with the animal will happen on kilometres $x$ to $x + \Delta x$ on the line $l$ at a given time of the day (time range $t$ to $t + \Delta t$ in particular) given $\tau$ part of the year (e.g. month). If the estimated probability exceeds some threshold value, a warning for the driver will be raised.
We use the following parameters of the model:
\begin{itemize}
    \item $\mu(\tau)$ -- the expected number of accidents in particular day, in  month denoted by $\tau$; formally $\mu(\tau)  = \sum_k k p(k, \tau)$ where $k$ numerates number of accidents in one day, and $p(k, \tau)$ is probability that $k$ accidents would occur in particular day in month $\tau$; however in our case even if $k > 1$ various accidents usually occurs at different time intervals and multiple daily accidents are assumed to be independent;  
    \item $p\left(t, \Delta t | \tau \right)$ -- the conditional probability that given the 
    accident occurs at the particular day of month $\tau$, it occurs in the time interval  $t, t+\Delta t$;   
    \item $p(l | \tau ,t, \Delta t )$ -- the conditional probability, that given the accident occurs at the particular day of month $\tau$, and in the time interval $t+\Delta t$ it occurs on the line $l$;
    \item $p\left(x, \Delta x | \tau, t, \Delta t , l\right)$ -- the conditional probability that given the accident occurs at the particular day of month $\tau$, in the time interval $t+\Delta t$ and on the line $l$, it occurs between $x$ and $x + \Delta x$ kilometre posts (or milepost) of this line.
\end{itemize}

Assuming that accidents are independent, the probability that the accident will occur at the particular day of month $\tau$, in time interval $t, t + \Delta t$ (of this day), on the line number $l$, between kilometre posts $x$ and $x + \Delta x$ reads
\begin{equation}\label{eq:probability}
\begin{split}
    &p\left(\tau, t, \Delta t , l, x, \Delta x\right) = p\left(x, x + \Delta x | \tau, t, \Delta t, l\right) p\left(l | \tau ,t, \Delta t \right) p\left(t,\Delta t | \tau \right) \mu(\tau).
    \end{split}
\end{equation}

Here, because of the limited data sample, we do not distinguish between days of the month. This is obviously the simplification, but still allows us to reach meaningful results. Additionally, we take advantage of the observation that the daily profile is dependent on the particular season of the year (i.e. summer or winter), and it can be determined by months. Such a dependency can be observed in Fig.~\ref{fig:time_profile}. Furthermore, we assume particular sizes of the time window $\Delta t$ and space window $\Delta x$ in the spatial-temporal model of accidents. 

One should note that in our model, the independence of spatial and temporal profiles of accidents is assumed. This would hold in general as long as the migration paths of the animals did not have significant seasonal variations. We do not expect this in our regional approach. However, one should consider this point in the case of adapting the method to the particular case. Thanks to the assumption of independence of spatial and temporal profiles, we split our model into the spatial part
\begin{equation}
 p(l, x, \Delta x) = p\left(x, \Delta x | l, \tau, t, \Delta t\right) p(l | \tau, t, \Delta t ) = p(x, \Delta x | l) p(l)
 \label{eq:xprofile}
\end{equation}
and the temporal part
\begin{equation}
 p(\tau, t, \Delta t) = p(t, \Delta t | \tau) \mu(\tau).
 \label{eq:tprofile}
\end{equation}

Considering the above considerations, the probability that the particular train on the particular day takes part in the accident reads
\begin{equation}
    p_{pt}\left(\tau, t, \Delta t, l, x,  \Delta x\right) = \frac{p\left(\tau, t,\Delta t, l, x, \Delta x\right)}{m\left(\tau, t, \Delta t, l, x, \Delta x\right)}  = 
   \frac{ p(\tau, t, \Delta t) p(l, x, \Delta x)}{m\left(\tau, t, \Delta t, l, x, \Delta x\right)}.
    \label{eq:probs}
\end{equation}
We denote by $m\left(\tau, t, \Delta t, l, x, \Delta x\right)$ the number of trains that passes the segment $x$ - $x + \Delta x$ of the line $l$ in the time interval $t$ to $t + \Delta t$ at the particular day of month $\tau$ . In the case where 
\begin{equation}
p_{pt}\left(\tau, t, \Delta t, l, x,  \Delta x\right) > p_{\text{threshold}}
\label{eq::p_threshols_warnings}
\end{equation}
a warning is raised, resulting in the given line segment being marked with a warning. The pseudocode describing the procedure for calculating the warnings for lines in a given period of time is provided in \textsc{BayesWarnAnimals} Procedure~\ref{proc:warnings}. Bear in mind that the procedure enables computation as well as updating of warning with new data sets.
\begin{algorithm}[H]
	\begin{algorithmic}
		\REQUIRE probability $p_{\text{threshold}}$, line $l$, month $\tau$, time period $[t,t+\Delta t)$
		\FOR{$x_i=x_0$ to $x_f$ step $\Delta x$} 
		\STATE $\mu(\tau)$ $\leftarrow$ calculated from data \COMMENT{Estimate $\mu$ for particular day of the given month $\tau$}
		\STATE $p(\tau, t, \Delta t)$ $\leftarrow$ calculated from data \COMMENT{Estimate $p$ for the particular day of the given month $\tau$ and time of the day range $[t, t+\Delta t)$}
        \STATE $p(l)$ $\leftarrow$  calculated from data \COMMENT{Estimate $p$ for given line $l$}
        \STATE $p(l, x, \Delta x)$ $\leftarrow$  calculated from data \COMMENT{Estimate $p$ for the given kilometre posts on line $l$}
        \STATE $m\left(\tau, t, \Delta t, l, x, \Delta x\right)$ $\leftarrow$ from timetable \COMMENT{Number of trains at particular kilometre posts of line $l$ at the particular day in month $\tau$ in time interval $[t, t+\Delta t)$}
        \STATE $p_{pt}(\tau, t, \Delta t, x_i, \Delta x)$ $\leftarrow$ Eq.~\ref{eq:probs}
		\IF{$p_{pt} > p_{\text{threshold}}$} 
		\STATE $w(x_i) \leftarrow$ 1
		\ELSE
		\STATE $w(x_i) \leftarrow$ 0
		\ENDIF   
		\ENDFOR
	\end{algorithmic}
	\caption{\textsc{BayesWarnAnimals} for updating the warnings for the selected line $l$ at month $\tau$, during the time period $[t,t+\Delta t)$. The procedure requires initial estimates for conditional probabilities, which could be updated with new data about the collisions.}
	\label{proc:warnings}
\end{algorithm}

To demonstrate the application of the model for realistic data, in the next section, we provide an example of data-driven analysis based on the data on animal-train accidents of the period of three years. 

\section{Data analysis}\label{sec:data_analysis}

The data used in the following analysis have been supplied by the regional railway operator operating in the Silesia region in the southern part of Poland. The region is among the heavily urbanised regions in Central Europe, but at the same time with rich wildlife. The operator uses the railway network that belongs to the national network operator (PKP PLK S.A.) and is independent of the rail operator by national law. We use the line numbering set by the network operator.
\subsection{Data description}

The railway network studied in the presented analysis runs through different areas. A significant part runs through the urban areas, in particular in Silesia Agglomeration, including cities such as Katowice, Sosnowiec, Mysłowice, Tychy, Zabrze, and Gliwice. However, the network also partially covers rural areas, and, what is the most important from a research point of view, it also partially covers forest areas with intense wild life. It should be stressed that this part of the network is not equipped by the network operator with any warning devices, such as the UOZ (animal protection device acoustic deflector)~\cite{UOZ-1device}. The trains are also not equipped with any animal warning devices. 

In the last period of timetabling (that is, from October to December 2022), the operator served $415$ trains per day. The traffic volume on individual sections of the railway lines was analysed based on the actual railway operator's schedule from the above-mentioned period. However, one should note that the distribution of the number of trains is not uniform in the day profile. For example, between 12 p.m. and 3 a.m., the operator serves almost no trains. The highest density of trains is during rush hours, that is, 6 a.m. - 8 a.m. and 3 p.m. - 5 p.m. in each of these windows, there are approximately $20\%$ trains on the network. Hence, we have introduced the non-uniform daily profile of number of trains.

Finally, the analysis time period partially overlaps with the COVID-19 pandemic period, where the number of trains was significantly reduced during the lockdown months. The railway operator reduced the number of trains by about $16\%$ in the period from January to August $2020$, then the transport returned to normal. We have not taken into account this unusual reduction of number of trains in our model (as an unusual event, it is not likely to be repeated in future). Despite this simplification, we still have reached the full meaning of the results.

\begin{figure}[t!]
\centering
\subfigure[]{\includegraphics[width=0.45\textwidth]{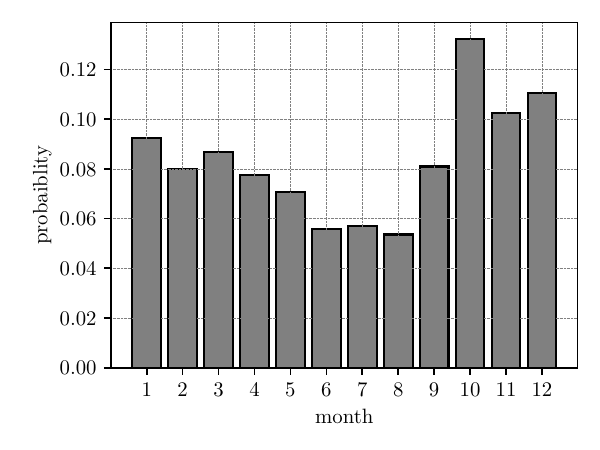}\label{fig:time_profile:monthly}}
\subfigure[]{\includegraphics[width=0.45\textwidth]{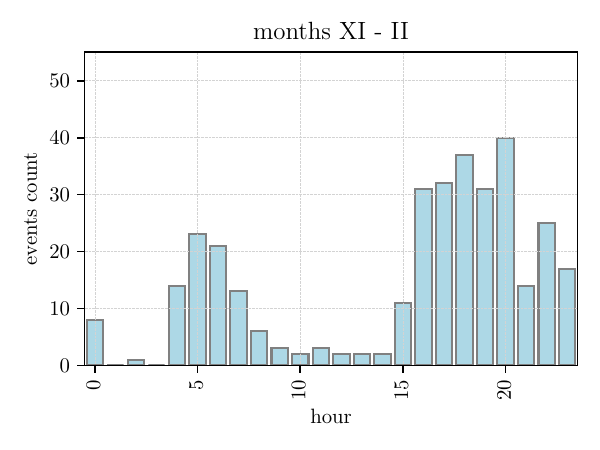}\label{fig:time_profile:XI_II}}
\subfigure[]{\includegraphics[width=0.45\textwidth]{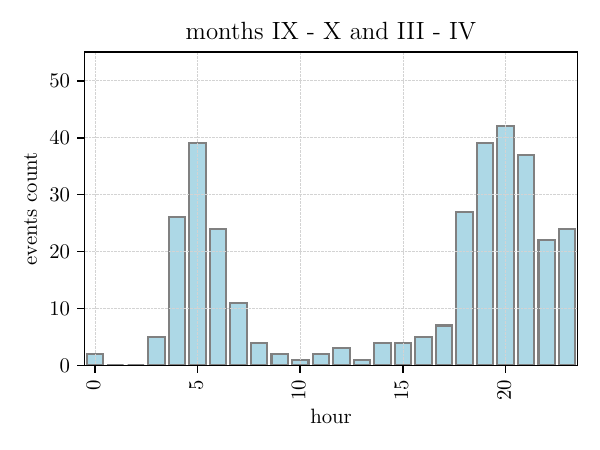}\label{fig:time_profile:IX_X_III_IV}}
\subfigure[]{\includegraphics[width=0.45\textwidth]{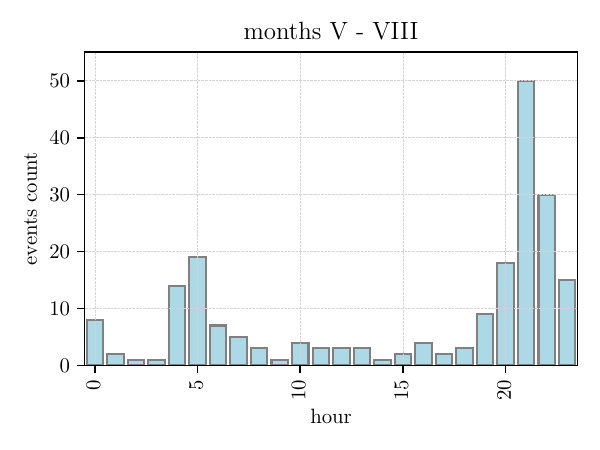}\label{fig:time_profile:V_VIII}}
\caption{Yearly (\subref{fig:time_profile:monthly}) and daily (\subref{fig:time_profile:XI_II}-\subref{fig:time_profile:V_VIII}) profiles of accidents involving wild animals. Data collected in the period from 2020 to 2022.}
\label{fig:time_profile}
\end{figure}

\begin{figure}[t!]
\centering
\subfigure[]{\includegraphics[width=0.32\textwidth]{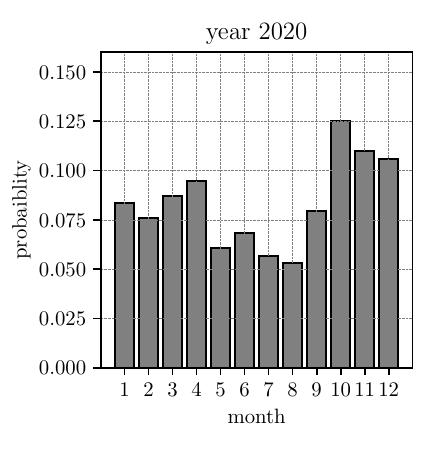}\label{fig:time_profile:monthly2020}}
\subfigure[]{\includegraphics[width=0.32\textwidth]{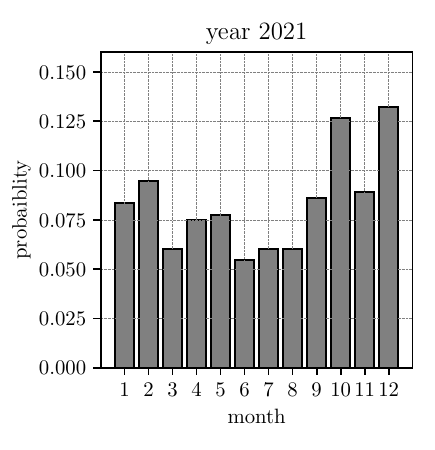}\label{fig:time_profile:monthly2021}}
\subfigure[]{\includegraphics[width=0.32\textwidth]{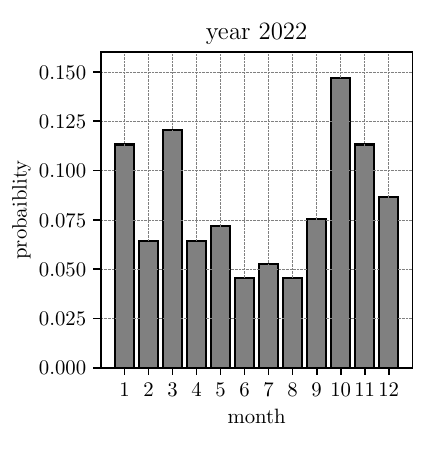}\label{fig:time_profile:monthly2022}}
\caption{Monthly profiles of accidents involving wild animals over particular years. We observe similar profile for all years. To demonstrate the stability of model, one can compare 2020~\subref{fig:time_profile:monthly2020} and 2021~\subref{fig:time_profile:monthly2021} as in the first half of 2020  train traffic was reduced due to COVID-19 lockdown.}
\label{fig:time_profile_y}
\end{figure}

\begin{figure}[t!]
\centering
\includegraphics[width=0.45\textwidth]{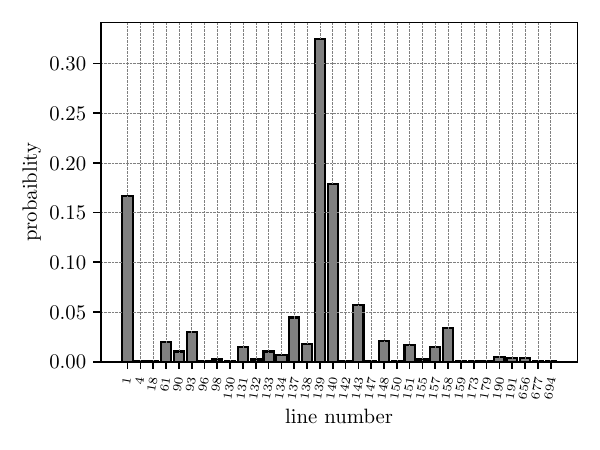}
\includegraphics[width=0.45\textwidth]{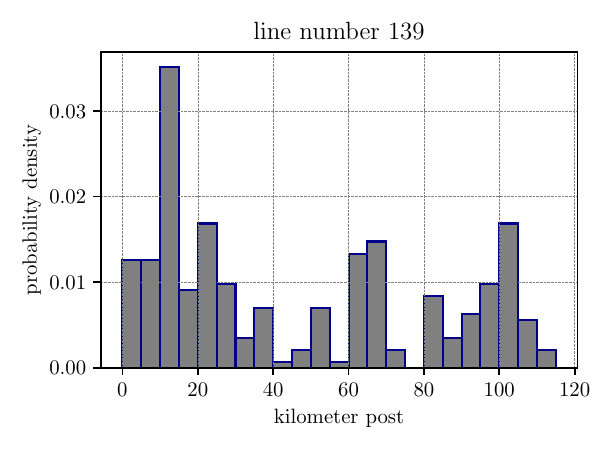}
\caption{Probability of accidents with animals for various lines, and spatial location, data collected in 2020 to 2022. }\label{fig:accidents_lines}
\end{figure}

The above-mentioned railway operator uses light electric multiple units (EMUs) that may be vulnerable to collisions with animals. Data on animal accidents are collected from drivers reports. Drivers are obliged, by the operator's regulations, to report any distortions and unusual events during the service of the train. Each such accident is annotated by the following set of parameters: date (day, month, year), time, number of line, kilometre post of the line, and the species. Data cover $3$ years in the period from 2020 to 2022. Additionally, we used a set of testing data from the first 4 months of 2023. From the 2020-2022 period, the profile of accidents has been demonstrated in Fig.~\ref{fig:time_profile} (time profiles) Fig.~\ref{fig:time_profile_y} (detailed monthly profiles) and Fig.~\ref{fig:accidents_lines} (spatial profiles).

One should observe that daily profiles in Fig.~\ref{fig:time_profile} are not symmetric in reference to early daylight (drawn) and late daylight (dusk), we observe more accidents in the dusk case. From Fig.~\ref{fig:accidents_lines}, most accidents have been recorded for line numbers $1$, $139$ and $140$. Hence, we will pay special attention to these lines and they will serve us to demonstrate the results of our model. 

In Table~\ref{tab::monthly} we present monthly statistics of train collision data to account for the size of the data set and its variability on a yearly basis. Referring to Table~\ref{tab::monthly} and to profiles in Fig.~\ref{fig:time_profile_y} we can conclude no meaningful differences between years. This demonstrates the robustness of the method, regardless of the COVID-10 lockdown in the first half of 2020. 

\begin{table}[]
\begin{tabular}{|l|l|llll|}
\hline
     & yearly & \multicolumn{4}{l|}{monthly}    \\ \hline
year & total  & \multicolumn{1}{l|}{min.} & \multicolumn{1}{l|}{max.} & \multicolumn{1}{l|}{mean} & std  \\ \hline
2020 & 264    & \multicolumn{1}{l|}{14}      & \multicolumn{1}{l|}{33}      & \multicolumn{1}{l|}{22.0} & 5.67 \\ \hline
2021 & 348    & \multicolumn{1}{l|}{19}      & \multicolumn{1}{l|}{46}      & \multicolumn{1}{l|}{29.0} & 8.37 \\ \hline
2022 & 265    & \multicolumn{1}{l|}{12}      & \multicolumn{1}{l|}{39}      & \multicolumn{1}{l|}{22.1} & 8.39 \\ \hline
\end{tabular}
\caption{Monthly statistics of number of accidents. Observe, that yearly means are similar, which reflects the statistical stability of data.}\label{tab::monthly}
\end{table}

\subsection{Profile estimation}

As mentioned before, to be independent of the particular theoretical profiles tied to particular spices, we estimate profiles from raw data only. This is an important feature of the proposed approach, as it ensures that the model will not be biased by the initial choice. 

Let $T$ be the number of days, data are collected, and $N$ the number of accidents then. Let $\tau$ enumerate periods of the seasonal profile, month in our case (we then have $\tau_{\text{total}} = 12$ monthly profiles). Based on this, we estimate $\mu(\tau)$ by
\begin{equation}
    \mu(\tau) = \frac{N_{\tau}}{T/ \tau_{\text{total}}}.
    \label{eq::nu}
\end{equation}
Here, $N_{\tau}$ is the number of accidents that appear in a month $\tau$, and $T/ \tau_{\text{total}}$ is (approximated) the number of days in a particular month from data (for example if $T = 365$ then $T/ \tau_{\text{total}} \approx 30$).

To estimate $p(\tau, t, \Delta t)$, we group months into seasons determined by $\mathcal{T}_i$, and check the season for temporal profiles. Here, for the sake of simplicity, we use three seasons, namely:
\begin{enumerate}
\item Short daylight case (late autumn, winter) -- months: November, December, January, February $\mathcal{T}_1 \in \{XI, XII, I, II\}$;
\item Long daylight case, months: May, June, July, August $\mathcal{T}_2 \in \{V, VI, VII, VIII\}$;
\item Intermediate daylight case, months: March, April, September, October $\mathcal{T}_3 \in \{III, IV, IX, X\}$.
\end{enumerate}
However, one should note that this assumption can be lifted, and generalisation is possible.

We denote by $N^{(\mathcal{T}_i)}$  the number of accidents recorded in $i$'th such season, and by $N^{(\mathcal{T}_i)}_{t, t + \Delta t}$ the number of accidents recorded in $i$'th season in the hour interval $t$ to $t + \Delta t$. Then we use the following estimate:
\begin{equation}
    p(\tau, t, \Delta t) = \frac{N^{(\mathcal{T}_i)}_{t, t + \Delta t}}{N^{(\mathcal{T}_i)}} \text{ chose $i$ such that }  \tau \in \mathcal{T}_i.
    \label{eq::p_tau_t}
\end{equation}

Let us now take advantage of the priori-discussed assumption that temporal and spatial profiles are independent. Then, we can estimate $p(l)$ as
\begin{equation}
    p(l) = \frac{N_l}{N},
    \label{eq::p_l}
\end{equation}
where $N_l$ is the number of accidents on the line $l$.
Finally, we estimate $p(l,x, \Delta x)$ by
\begin{equation}
    p(l, x, \Delta x) = \frac{N^{(l)}_{x, x + \Delta x}}{N_l},
    \label{eq::p_l_x}
\end{equation}
where $N^{(l)}_{x, x + \Delta x}$ is the number of accidents on the line $l$ between kilometre post $x$ and $x + \Delta x$. 

\begin{figure}
    \centering
\includegraphics[width=0.32\textwidth]{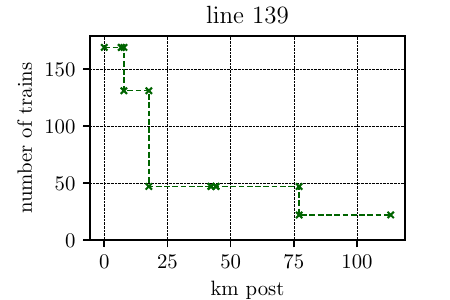}
\includegraphics[width=0.32\textwidth]{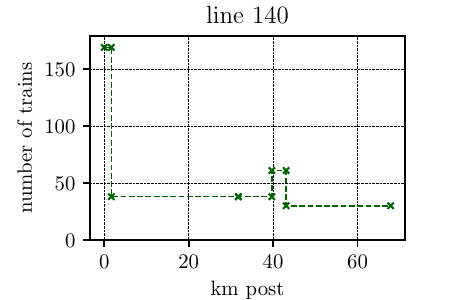}
\includegraphics[width=0.32\textwidth]{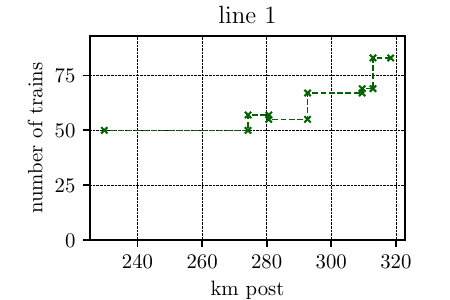}
\caption{Number of trains during the working days for the selected lines. One should note that lines $139$ and $140$ shares their starting points. Additionally, one can note various profiles of number of trains for various lines. This suggests the number of trains should be taken implicitly into our model.}
\label{fig:number of trains}
\end{figure}
Referring to Fig.~\ref{fig:number of trains}, the number of trains varies within the line and between lines. Hence, we need to take into account the number of trains in our model. In practice, this quantity is given by the railway operator, or can be read out from the public timetable. The number of trains at a particular line, and between particular km posts, is denoted as $m(l,x, \Delta x)$. Here, we assume no seasonal variation as well no day to day variation, that is the approximation adopted in our research. Thus we have, 
\begin{equation}
    m(t, \Delta t, l, x, \Delta x) = m(\tau, t, \Delta t, l, x, \Delta x) = m(l,x, \Delta x) \alpha(t, \Delta t)
    \label{eq::m_t_l}
\end{equation}
where $\alpha(t, \Delta t)$ is the temporal profile considering peak hours and off-peak hours and night break
\begin{equation}
    \alpha(t, \Delta t) = \begin{cases} 0.05 / (4 \Delta t) &\text{ if } 0 \leq t < 4 \\
    0.4 / (6 \Delta t) &\text{ if } 6 \leq t < 9 \vee 15 \leq t < 18 \\
    0.55 / (14 \Delta t) &\text{ elsewhere } 
    \end{cases}.
    \label{eq::alpha}
\end{equation}
It is derived from the observation that in peak hours, i.e. 6 a.m. - 9 a.m. and 3 p.m. - 5 p.m., $40 \%$ of trains are served. Further, it is assumed that in night hours, 12 a.m. - 3 p.m., only $5 \%$ of trains are served. The rest of the trains are served during off-peak hours. Finally, one can use the formula provided in Eq.~\eqref{eq:probs} to compute the probability of the accident of the particular train on a particular day given $\tau$, $t$, $\Delta t$, $l$, $x$, $\Delta x$.
\subsection{Illustrative example}

In this example, we present step-by-step computation of the warning that can be applied with almost no IT support if necessary. Data for this example are presented in Tab.~\ref{tab:example}. We have $N = 877$ accidents recorded in 2020-2022, that is, $3$ years ($T = 3 \cdot 365$).
\begin{table}[h!]
\begin{tabular}{|l|l|l|l|l|l|l|}
\hline
\begin{tabular}[c]{@{}l@{}}  N\end{tabular} & \begin{tabular}[c]{@{}l@{}}\# acc. \\ I (Jan) \\$N_{\tau=I}$\end{tabular} & \begin{tabular}[c]{@{}l@{}} \#acc. season\\ XI, XII, I, II \\$N^{(\mathcal{T}_1)}$ \end{tabular} & 
\begin{tabular}[c]{@{}l@{}} \#acc. at  6 p.m. \\ in XI, XII, I, II \\$N^{(\mathcal{T}_1)}_{18, 18 + 1}$ \end{tabular}& \begin{tabular}[c]{@{}l@{}} \# acc. \\ l. $139$ \\ $N_{l=139}$\end{tabular} & \begin{tabular}[c]{@{}l@{}} \# acc.  l. 139  \\  km $10-15$ \\$N^{(l = 139)}_{10, 10+5}$\end{tabular} & \begin{tabular}[c]{@{}l@{}}\# trains l. $139$ \\ km $10-15$ \\ $m(l=139, 10, 15)$\end{tabular} \\ \hline
877  & 81 & 338 & 37  & 285  & 50  & 131  \\ \hline
\end{tabular}\caption{Data for illustrative example}\label{tab:example}
\end{table}

Using the date specified in Table~\ref{tab:example}, and executing the steps described in Procedure~\ref{proc:warnings}, the warnings are calculated as follows: 
\begin{enumerate}
    \item Estimate $\mu(\tau)$ using Eq.~\eqref{eq::nu}.\\
    For January ($I$) we had $81$ records, yielding $\mu(\tau) \approx 0.89$.
    \item Estimate $p(\tau, t, \Delta t)$ in season $\{XI, XII, I, II \}$, and $t = 18$h and $\Delta t = 1$h.\\
    From Eq.~\eqref{eq::p_tau_t} we have $p(t, \Delta t | \tau) \approx 0.11$.
    \item Estimate $p(l)$, for  line $139$.\\
    From Eq.~\eqref{eq::p_l} have $p(l) \approx 0.33$.
    \item Estimate $p(l, x, \Delta x)$, for $x = 10$km and $\Delta x = 5$km.\\
    From Eq.~\eqref{eq::p_l_x} we have $p(x, \Delta x | l) \approx 0.175$.
    \item From Eq.~\eqref{eq:xprofile} $p(x, \Delta x, l) \approx 0.056$, while from Eq.~\eqref{eq:tprofile} we have $p(\tau,t,\Delta t) \approx 0.1$.
    \item On the line $139$ between $10$ and $15$ km post there are $131$ trains per day.\\
    Using Eqs.~\eqref{eq::m_t_l}, \eqref{eq::alpha} at the off-peak, we have $m(\tau,t,\Delta t, l, x, \Delta x) = 131\cdot 0.55/14 \approx 5.15$.
    \item From Eq.~\eqref{eq:probs}, we obtain $p_{pt}\left(\tau, t, \Delta t, l, x,  \Delta x\right) \approx 0.0011$.
    \item Given $p_{\text{threshold}} = 0.001$ according to Eq.~\eqref{eq::p_threshols_warnings} the warning is raised, i.e. $w(x_i = 10) = 1$.
\end{enumerate}

The above calculation provides only an illustrative example of the proposed method. More comprehensive results, involving computations for several lines, are presented in the next subsection.

\subsection{Evaluation results}

\begin{figure}    \includegraphics[width=0.38\textwidth]{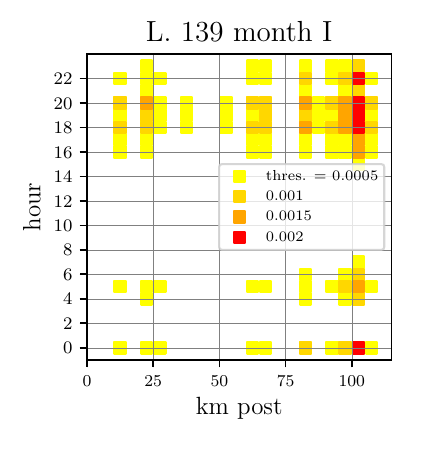}
\includegraphics[width=0.23\textwidth]{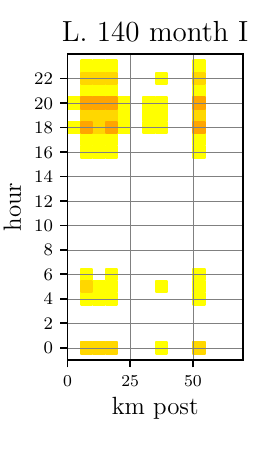}
\includegraphics[width=0.32\textwidth]{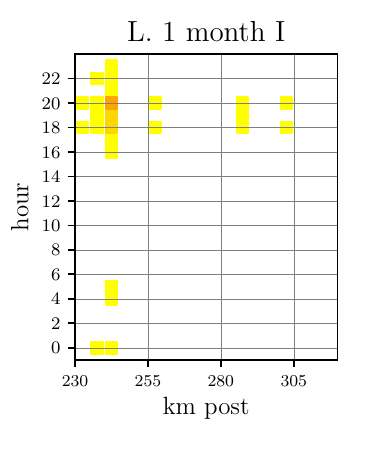}
\includegraphics[width=0.38\textwidth]{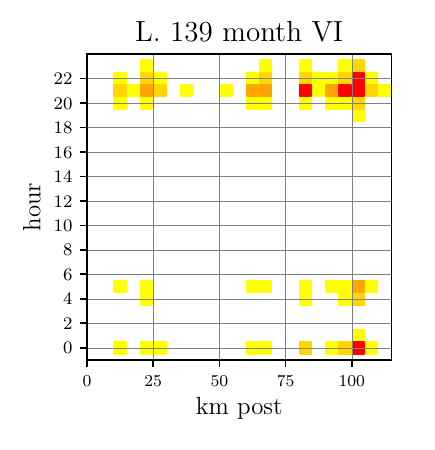}
\includegraphics[width=0.23\textwidth]{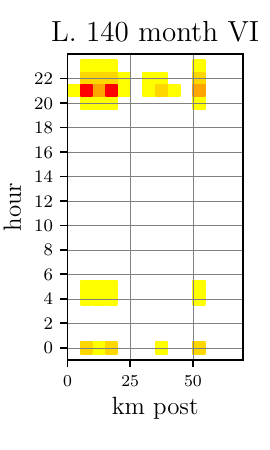}
\includegraphics[width=0.32\textwidth]{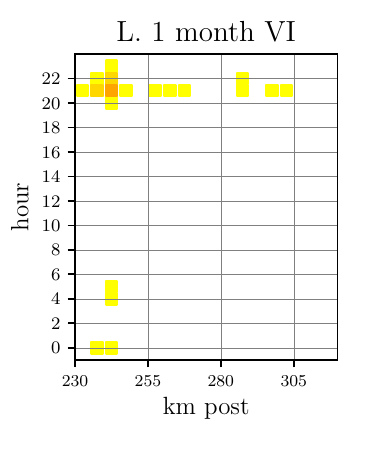}
\includegraphics[width=0.49\textwidth]{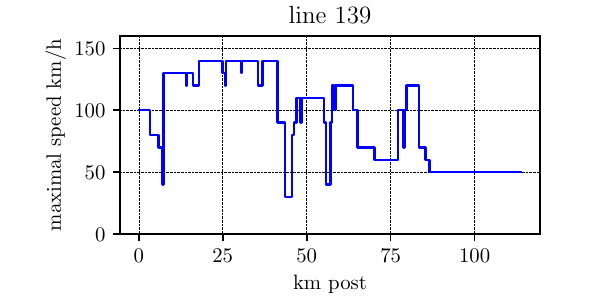}
\includegraphics[width=0.49\textwidth]{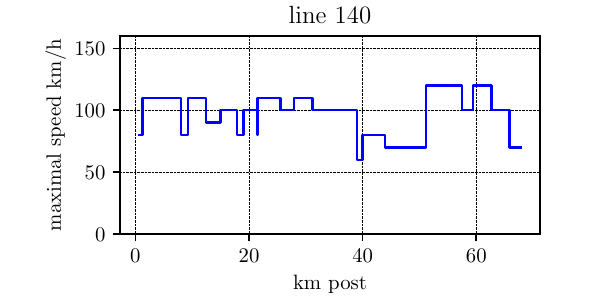}
\caption{Warnings for various locations and times, for most susceptible lines, and maximal speed profiles on chosen lines (bottom panel) from official documents of PKP PLK S.A.network operation~\cite{regulamin_PLK}. Yellow - orange - red rectangles represent warnings, and their colours represent the thresholds value $p_{\text{threshold}}$, see Eq.~\eqref{eq::p_threshols_warnings}. One should note, the end of the line $139$ with low speed and a high number of warnings.}\label{fig:warnings_various_lines}
\end{figure}
\begin{figure}
\includegraphics[width=0.5\textwidth]{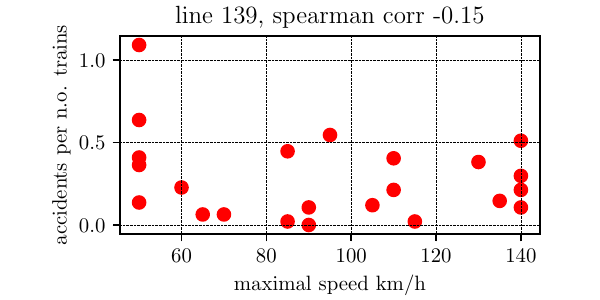}
\includegraphics[width=0.5\textwidth]{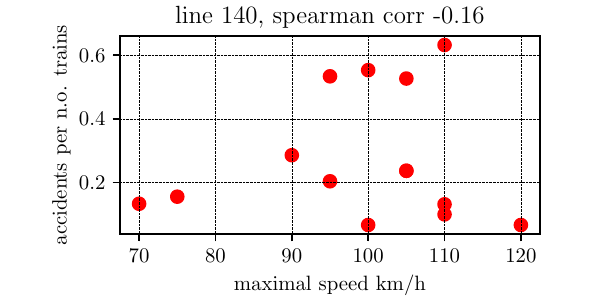}
\caption{Number of accidents in $\Delta_x = 5$ km ranges per train vs maximal speed on the line, for chosen lines.  Observe no positive correlation between speed and the number of accidents.}\label{fig:speed_corr}
\end{figure}

For a more thorough warning analysis, the most susceptible lines of Fig.~\ref{fig:accidents_lines} (left panel) were chosen. Maps of warning computed by \textsc{BayesWarnAnimals} Procedure~\ref{proc:warnings} are presented in Fig.~\ref{fig:warnings_various_lines} (winter time upper panel and summer time middle panel). 
Observe that in summer time there is a wide gap (between 6 a.m. and 7 p.m. where warnings are not raised). This approves the general findings that accidents with wild animals are less probable in daylight~\cite{BHARDWAJ2022114992}.

Observe also that there are many warnings on the line $139$ beyond $80$ km post. This is because there are a high number of accidents, given the small number of trains, see Fig.~\ref{fig:number of trains}. For the vehicle-based animal warning system, the number of trains on a given part of the railroad has to be taken into account. As an extreme example, let us imagine the situation where accidents are frequent with a very low number of trains. Then the probability of an accident for the particular train is high. 

To follow this line of analysis, one can note that the results in Fig.~\ref{fig:warnings_various_lines} do not fully reflect hot-spots of a large number of accidents, in Fig.~\ref{fig:locations}. Hence, we state the open question, then, what impact on the number of accidents would have a (hypothetical) increase in the number of trains at the end of line 139, to values such as at the beginning of line 139; see Fig.~\ref{fig:number of trains}.

The large number of accidents above $80$ km post-line 139 appeared regardless of low speed in this fragment of line. This observation is contrary to the observation in~\cite{Roy2017} for elephants, where train speed was recognised as one of the main factors in animal accidents. At this part of line $139$ train speed is limited (approximately $50$ km/h; see lower panel of Fig.~\ref{fig:warnings_various_lines}). To evaluate this important observation, in Fig.~\ref{fig:speed_corr} we demonstrate the relation between train speed and the number of accidents per train for chosen lines. What is important is that we have not observed the positive correlation between these two quantities. This is most likely because many accidents occur in low-speed forest areas, where railway traffic is not particularly dense. Therefore, in this area, animals are not scared away from the track by fast-moving trains. This important observation should be tested on more of railway systems, where accidents with animals like roe deer or wild boar are recorded.

\begin{figure}[ht!]
\includegraphics[width=0.37\textwidth]{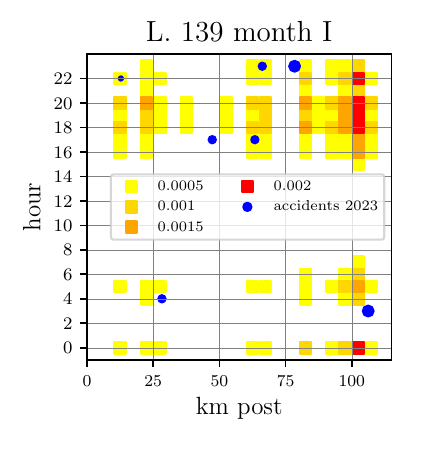}
\includegraphics[width=0.37\textwidth]{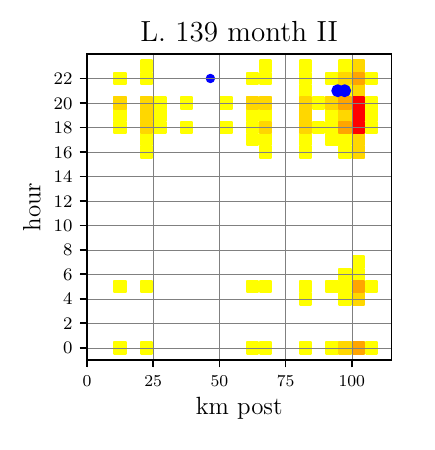}
\includegraphics[width=0.23\textwidth]{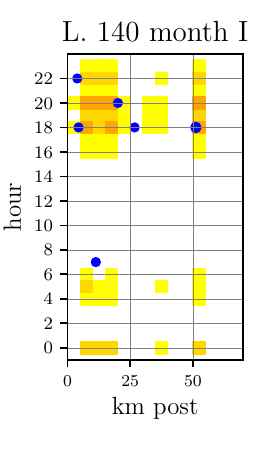}
\includegraphics[width=0.37\textwidth]{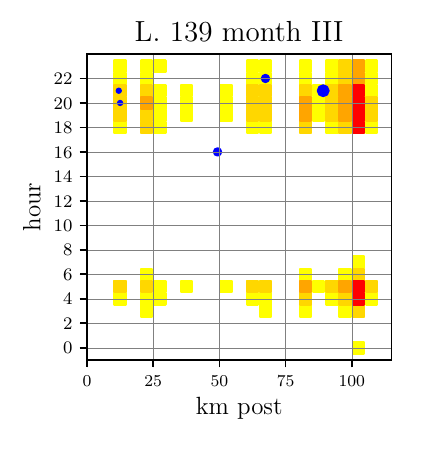}
\includegraphics[width=0.37\textwidth]{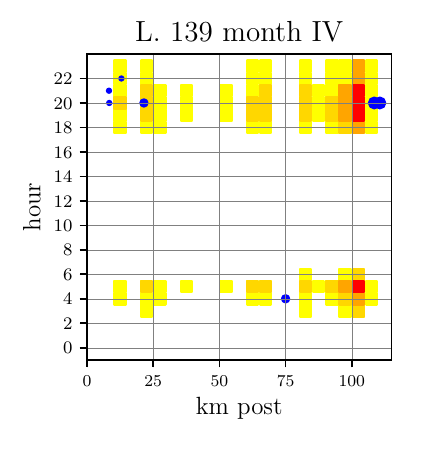}
\includegraphics[width=0.23\textwidth]{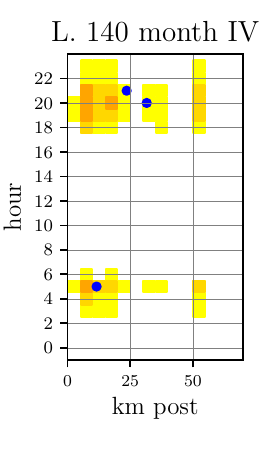}
\caption{Warning computed chosen lines from years 2020-2022 (yellow - orange - red rectangles with colours representing the thresholds value $p_{\text{threshold}}$) and the actual number of accidents in the first four months of 2023 (blue dots). The size of blue dots is scaled inversely to the number of trains in the location where accidents were recorded, see Fig.~\ref{fig:number of trains}. Observe, that most of new accidents occur in locations (or close to locations) where warnings computed from past data were raised. Furthermore, most collisions occur in late afternoon, early evening - the time of dusk and early night.}\label{fig:warns_and_new_dats}
\end{figure}

\begin{figure}[ht!]
\includegraphics[width=\textwidth]{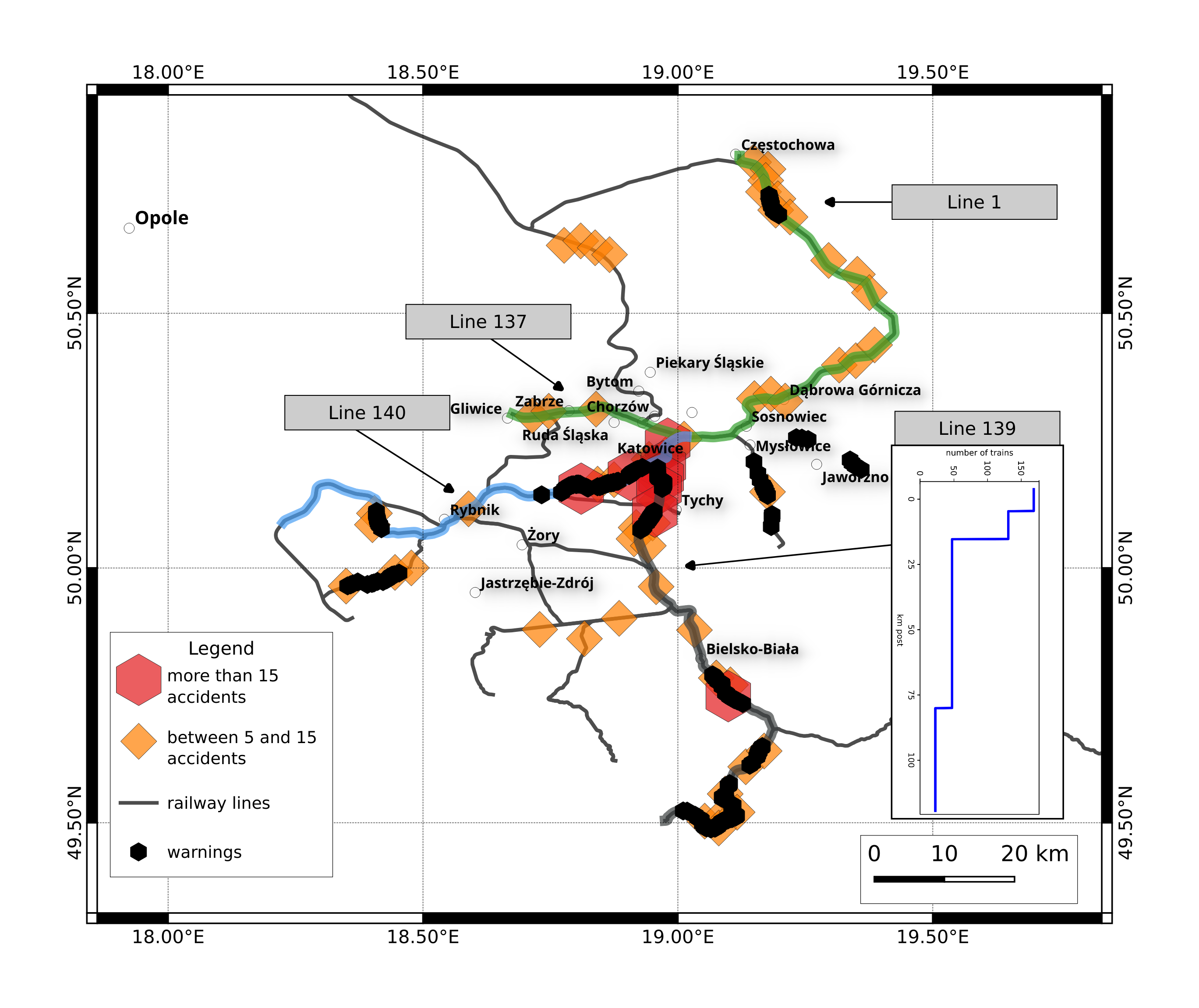}   \caption{Demonstrative view of warnings inferred from the statistical analysis of the number of accidents and the traffic volume for the period of three years (warnings for January 6 p.m., threshold value $p_{\text{threshold}} = 0.001$). The results are based on the all-year information about the number of accidents. Black spots represent lines with the increased probability of accidents. See Fig.~\ref{fig:warnings_various_lines} for the numerical values for the corresponding lines.}\label{fig:map_of_wanrs}
\end{figure}

In the above paragraphs, we have presented the descriptive features of our model. Let us now move to its predictive features. For this reason, we use the previously mentioned data from years 2020-2022 as the training set. New data from the first 4 months of 2023 are used as the test set. In Fig.~\ref{fig:warns_and_new_dats} it is shown that most new accidents (testing set data are denoted by blue dots) occur in places or close to places for which warnings (computed from training set) were raised. This demonstrates the stability of the model, regardless of disruptions in training data (such as data from the COVID pandemic lockdown, which were not filtered out from training data). It is also worth mentioning that we have achieved meaning-full patterns of collisions given available data from a short period of time (i.e. 3 years).

To demonstrate the spatial locations of accidents and warnings from the training data, see Fig.~\ref{fig:map_of_wanrs}. Accidents are from all data, while warnings are for a selected time and season (winter) and threshold parameter $p_{\text{threshold}} = 0.001$. From these data, at least three potential animal habitat regions can be deduced. As already mentioned, we have interesting results at the end of the line $139$ (the bottom of the figure~\ref{fig:map_of_wanrs}) as a high number of accidents coincides with a relatively low number of trains on this fragment of line.

To present metrics of the model's performance, we opt for the accuracy measure computed as the $p_{\text{threshold}}$ dependent fraction of accidents from the testing set that fall into warnings computed via Procedure~\ref{proc:warnings} from the training set. This accuracy is compared with the $p_{\text{threshold}}$ dependent percentage of warnings, namely: the actual number of warnings over a maximal possible number of warnings. The maximal possible number of warnings is the number of warnings at $p_{\text{threshold}} = 0$ (indeed a little below $0$ as we have a sharp inequality in Eq.~\eqref{eq::p_threshols_warnings}).

\begin{figure}[t!]
\centering
\subfigure[]{\includegraphics[width=0.45\textwidth]{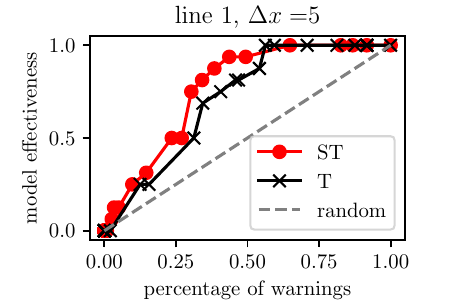}\label{fig:ROC[1]5}}
\subfigure[]{\includegraphics[width=0.45\textwidth]{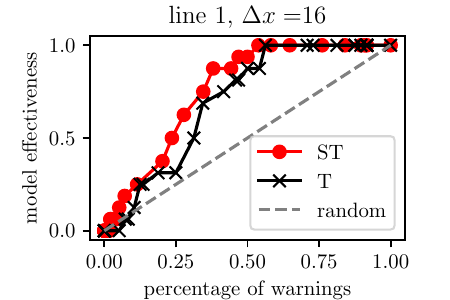}\label{roc:[1]16}}
\subfigure[]{\includegraphics[width=0.45\textwidth]{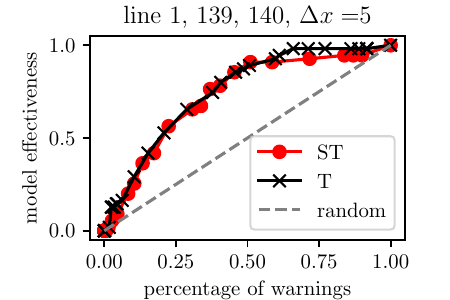}\label{fig:ROC[1,139,140]5}}
\subfigure[]{\includegraphics[width=0.45\textwidth]{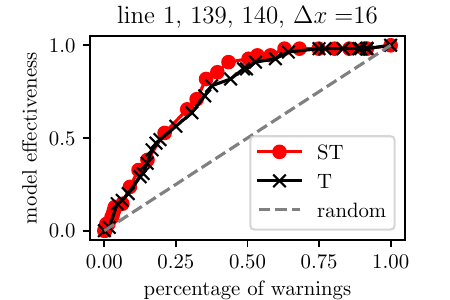}\label{roc:[1,139,140]16}}  \caption{Metrics of model effectiveness (measured as the percentage of accidents from testing data set that fits into warnings computed from the training set), for various lines and various $\Delta x$ parameter in Procedure~\ref{proc:warnings}. 
ST - original spatial and temporal testing scenario, T - temporal only testing scenario, where spatial dependency was not considered during testing.
Observe, that for line number $1$ the ST testing scenario returns better results. }\label{fig::ROC}
\end{figure}

For the comprehensive analysis we compare our method with the one reviewed form~\cite{Poland-analysis} (regarded earlier as closest to our approach). In~\cite{Poland-analysis} only temporal (day-time and year-time) dependency of accidents was considered, disregarding the spatial dependency. Henceforth, our goal is to assess the impact of spatial dependency on the model's effectiveness.
 Henceforth, we use two different testing scenarios.
\begin{enumerate}
    \item Bayesian, the spatial and temporal - ST, where we check whether each accident (recorded in the testing set) fits into the spatial and temporal warning window computed by \emph{Procedure 1} from the training set. 
    \item Derived from~\cite{Poland-analysis}, only temporal- T, the whole line at a given time is considered to be under the warning if there appears the warning computed by \emph{Procedure 1}, at any location on the line at this time.
\end{enumerate}
Results of testing these scenarios, compared with random choice, are presented in Figure.~\ref{fig::ROC} - the plot was performed for various $p_{\text{threshold}}$ values ranging from $0$ to $0.01$.

The results in fig.~\ref{fig::ROC} (top panel) indicate that in the case of line number $1$ the spatial and temporal (ST) testing scenario returns better results, than the temporal (T) approach. This is the sound result as line number $1$ passes through variable landscapes (industrial, agricultural, and forests). This observation confirms the intuition that our approach is most effective for lines that pass varying types of landscape.

The results in Fig.~\ref{fig::ROC} (bottom panel) indicate similar results of ST and T testing scenario for aggregated data from lines number $1$, $139$, and $140$. Furthermore, the TS scenario slightly outperforms the T one for particular $\Delta x$ settings (Fig.~\ref{fig::ROC} (d)). From the train driver's point of view, the ST approach is more sound as it has the potential for hot-spot detection. This observation paves the path for the elaboration of new models on more precise hot-spot detection algorithms from the statistical data only. The important issue here is to tune model parameters such as spatial resolution $\Delta x$ in Procedure~\ref{proc:warnings}.

\section{Discussion and Conclusions}\label{sec:discussion}

In the presented work, we develop a method for supporting railway safety systems using statistical analysis of geospatial data on the history of accidents involving animals. The presented method is based on simple assumptions about the statistical independence of the probability distribution of events involving animals. Hence, no assumptions are required on the model of animal activity. Furthermore, the model can be implemented using the standard software available to railway management authorities and railway undertakings. Thus, the proposed method can be potentially used as long as historical data on accidents involving animals are available and can be associated with the railway topology. Concerning effectiveness, our model is best suited to lines that pass varying types of landscape.

One of the observations one can make from the data analysed in the presented work is that the large number of accidents is not directly related to the probability of an accident. Therefore, to provide the information to be included in a warning system, one needs to take into account the number of trains on the given line in the particular time period.
In the case of numerous accidents and a large number of trains, the probability of an accident per train may not be large. In this situation, infrastructure-based countermeasure would be better than train-based countermeasure. This suggests that to reduce the number of accidents involving animals, one should differentiate the types of countermeasures used. In particular,
\begin{itemize}
    \item train-based countermeasure should be used if the probability of the accident per train is large;
    \item infrastructure-based countermeasure, including animal protection device acoustic deflector, should be used if the number of accidents is significant, there are a lot of accidents, and many trains -- but probability of accidents per train may not be large.
\end{itemize}

Furthermore, we have used the daily pattern of accidents. The obtained results confirm that warnings should be issued late afternoon and early evening hours (dusk). It interacts with the observation of collisions with moose and roe deer in Sweden in~\cite{seiler2017wildlife}, where most collisions occur during dusk and dawn. Analogously, in~\cite{keken2017railway}, the analysis of accidents in Czechia reviles that most accidents occur at dawn, dusk or at night, since ungulate animals have greater physical activity at dusk and dawn~\cite{kuvsta2017effect}.
Our findings indicate that dusk time is the most risky in terms of accidents with trains and animals, which was also observed in other research in Poland, see e.g. Fig.~$2$ in ~\cite{Poland-analysis}. In our opinion, it is the important observation that may have some origins in railway traffic density (afternoon peak hours overlap with dusk time in winter, and night break of traffic coincides with night hours in summer) or some particular behaviour pattern of roe deers and wild boars.

Finally, we have also argued that a positive correlation between the speed of trains and the number of accidents is not always observed in the data set used in the presented study. There is even the suggestion of a negative correlation, as frequent and fast-moving trains may be scaring away animals from the track region. This contrasts with the results in \cite{visintin2018managing} for kangaroos. Therefore, the policy to reduce the number of accidents should be related to particular spices and particular environmental conditions. Again, as our method is not based on any assumptions about the animal activity, it can be easily tuned to any situation without developing any parametrical model.

Henceforth, knowledge of the behaviour of particular species of animals can be used in more sophisticated models of profiles. However, our model is simple and easily implementable and can be applied even in companies with limited IT support (which may be crucial in railway companies from less developed countries).

\paragraph{Acknowledgements}
The authors would like to thank Wojciech Gamon for making possible the cooperation with Koleje Śląskie sp. z o.o. and for helping to acquire the data set used in the presented study.

\end{document}